# Tensile behavior of stainless steel 304L to Ni-20Cr functionally graded material: experimental characterization and computational simulations

Lourdes D. Bobbio[1], Brandon Bocklund[1], Zi-Kui Liu[1], Allison M. Beese[1,2]*

[1]Department of Material Science and Engineering, Pennsylvania State University, University Park, PA 16802

[2]Department of Mechanical Engineering, Pennsylvania State University, University Park, PA 16802

* Corresponding author, email: amb961@psu.edu

**Abstract**

This study presents experimental and computational analyses of the phase composition and mechanical behavior of a functionally graded material (FGM) grading from 100 vol% 304L stainless steel (SS304L) to 100 vol% Ni-20Cr (NiCr) in 10 vol% increments. The mechanical behavior of the FGM was evaluated by extracting tensile specimens from SS304L-rich and NiCr-rich regions of the FGM, with the former behaving in a ductile manner and the latter being relatively brittle. The brittle response of the NiCr-rich sample was explained by the presence of the eutectic FCC+BCC phase in this sample, as identified by Scheil-Gulliver simulations along with experimental phase and composition analyses. The spatially varying yield and flow behavior of the FGM was determined through comparison of finite element analysis simulations with experimentally measured stress-strain curves and surface strain contours. The results highlight that the overall ductile behavior of SS304L and brittle behavior of NiCr are preserved in the SS304L/NiCr FGM but the formation of the brittle eutectic phase further reduced the ductility within the regions where it formed (90 vol% NiCr/10 vol% SS304L).

1. **Introduction**

Functionally graded materials (FGMs) fabricated by additive manufacturing (AM) can have compositions, microstructures, and therefore properties, that vary as a function of position. Conventional methods of joining dissimilar materials, for example, through welding, mechanical fasteners, or interlocks, result in abrupt transitions in properties (e.g., coefficient of thermal expansion, elastic moduli, strength), and fusion welding methods may be complicated by additional factors including lack of solubility or atomic structure mismatch. These factors can result in the formation of brittle phases and the development of residual stresses. Tailored gradients between dissimilar materials in FGMs may be used to mitigate these issues, while also allowing for multifunctional properties within a component.

FGMs can be used in place of monolithic components to reduce the cost of material and the final weight of the component. For example, Ni-based alloys may be joined with Ti alloys for applications where only certain regions of the component will encounter high-temperature environments [1]. Similarly, for components typically made of Ti-6Al-4V where the entire structure must maintain good corrosion resistance, but only part of the component sees high temperatures during use, a graded component of Ti-6Al-4V and stainless steel would maintain the corrosion resistance where needed, while reducing the overall cost [2]. For structural applications, the mechanical integrity across regions of varying composition within FGMs is critical. In the biomedical space, Ti-6Al-4V has long been used in joint replacements due to its strength, biocompatibility, and low density, but its high elastic modulus can result in stress shielding and bone resorption. Some studies have examined grading Mo to make a more optimal graded component to reduce the effect of this stress shielding [3]. The mechanical integrity would be of utmost importance given the load-bearing application of such a component. Other applications of FGMs include turbocharges and interchanges in the automobile industry [1,4,5], turbine blades and vanes in aircraft engines and gas turbines [1,6–8], propulsion system and airframe of spacecrafts [5,9], combustion chambers and nozzle liners in rocket engines [10], and nuclear pipelines [3,11].

Published studies on FGMs primarily evaluate microhardness when considering mechanical behavior; however, some studies have subjected FGMs to tensile testing across the gradient interfaces, as well. Hengsbach et al. investigated an FGM sample, fabricated by selective laser melting AM, which was primarily composed of SS316L with two small sections of H14 steel within the gauge region of the tensile specimen [12]. In addition to the FGM samples, the authors also fabricated monolithic samples of SS316L and H14 steel in order to compare with the FGM samples. In the graded sample, there was a sharp change in microhardness between the SS316L section (267 HV) and the H13 steel section (689 HV). While the FGM had a similar ultimate tensile strength (UTS) to the pure SS316L made by AM (645 MPa for the FGM, 560 MPa for SS316L, 1570 MPa for H13 steel), its elongation to failure of 20% fell between that of pure the pure H13 steel made by AM (5%) and the pure SS316L (42%). Measurements of strain fields during deformation on the FGM showed that the plastic deformation occurred solely in the SS316L section of the FGM, as the pure H13 sample had a significantly higher yield strength (1600 MPa) than the pure sample SS316L (550 MPa).

Kannan et al. examined the microstructure, tensile properties, and fatigue strength of a SS904L and Hastelloy C-286 FGM made by wire arc AM in order to evaluate the FGM for use in harsh environments [13]. Scanning electron microscope (SEM) analysis of the interface between the 100% SSS904L and 100% Hastelloy C-286 regions of the FGM showed good interfacial bonding with no cracking between the two materials. The interface was tested in tension, resulting in a UTS of 680 MPa, a yield strength of 311 MPa, and an elongation of 38%. The samples, which had a higher tensile strength than wrought SS904L (490 MPa), failed in the SS904L region of the sample.

Niendorf et al. examined the effect of changing processing parameters within a 316L stainless steel sample fabricated by laser powder bed fusion, in order to vary microstructure and mechanical properties within the single component [14]. The processing parameters were altered within the tensile specimen, with the bulk of the sample fabricated with a relatively low laser power of 400 W to produce small grains, and small regions within the gauge region fabricated at a relatively high laser power of 1000 W to produce larger grains, with all other processing parameters kept constant. Tensile testing confirmed that strain localized in the high power, coarse-grained sections of the tensile specimen, highlighting the feasibility of direct microstructure manipulation through the adoption of alternating processing parameters in PBF AM.

There is interest in combining Ni-20Cr with stainless steel alloys because of the superior high temperature strength and corrosion-resistance in aqueous environments of Ni-20Cr [15,16] and its slightly lower density compared to stainless steel (7.75 $g/cm^3$ for Ni-20Cr and 8 $g/cm^3$ for SS304L). An FGM of these two materials would reduce the overall component weight compared to a monolithic SS304L component, while maintaining the critical corrosion resistant properties of a monolithic SS304L component [17]. Additionally, there is an interest in using Ni-20Cr as an intermediate layer between stainless steel and titanium alloys to enable joining of stainless steel and titanium alloys without the formation of intermetallic phases in the direct joining of these two terminal alloys [2,18]. Sahasrabudhe et al. investigated a bimetallic structure of 410 stainless steel and Ti-6Al-4V fabricated using LENS, both with and without a Ni-20Cr intermediate layer [2]. The sample without the intermediate layer had numerous cracks resulting in delamination due to the formation of intermetallic phases. Ni-20Cr proved to be a good intermediate alloy, resulting in a crack-free, structurally sound final component. The authors stated that further analysis is necessary to determine the mechanical properties of these interfaces.

While Ni-20Cr is typically fabricated through traditional casting techniques [15], there have been recent studies investigating the use of AM to fabricate Ni-20Cr components. Song et al. [19] performed a parametric study that examined the microstructure, phase composition, and mechanical properties of Ni-20Cr components fabricated using laser powder bed fusion. The tensile strength of these AM components ranged between 318-365 MPa, which is comparable to conventionally processed Ni-20Cr (330 MPa).

The present study experimentally and computationally investigated the phases and properties of an FGM grading from 100 vol% 304L stainless steel (SS304L) to 100 vol% Ni-20Cr (henceforth referred to as NiCr) in 10 vol% increments. Mechanical tests spanning multiple gradient interfaces were used to study how the composition changes affected the mechanical properties of the FGM. Subsequent finite element analysis (FEA) simulations were performed to link the experimental measurements to the constitutive behavior of the individual gradient regions. The CALculation of PHAse Diagrams (CALPHAD) method was used to predict the phases that could form in the FGM during the AM process. Equilibrium solidification and Scheil solidification simulations were used to understand the phases that could form in each layer. Equilibrium isopleths were used to understand solid state phase transformations that occurred after solidification. Experimental composition and phase analyses of the two composition regions where the samples failed were performed and compared with the thermodynamic phase predictions at these compositions, as well as the full gradient path.

## 2. Methods

### 2.1. Experimental investigations

A pillar-shaped FGM sample grading from 100 vol% SS304L to 100 vol% NiCr was fabricated using a laser- and powder-based directed energy deposition (DED) additive manufacturing system (RPM 557 Laser Deposition System), and photograph of the sample is given in Figure 1a. Pre-alloyed NiCr powder (American Elements, -140/+325 mesh) and SS304L powder (Carpenter, -140/+325 mesh) were used. The FGM was fabricated such that the volume fraction of SS304L decreased, and the volume fraction of NiCr increased, by 10 vol% every 20 layers, resulting in 20 layers of each gradient composition as shown in Figure 2. The layer height was 0.38 mm. Deposition was performed with a YAG laser operated at 800 W with a 1.35 mm spot diameter, a laser scanning speed of 12.25 mm/s, hatch spacing of 0.735 mm, and hatch angle between subsequent layers of 45°. The deposition was performed in an Ar atmosphere to prevent oxidation.

The as-built pillar was removed from the baseplate and sectioned using wire electrical discharge machining (EDM). One half of the pillar was mounted and polished using standard metallographic grinding and polishing techniques with a final polish using a 0.05 µm silica suspension. An electrolytic oxalic acid etch was used to reveal microstructural features. This sample was used for elemental and phase characterization, as well as hardness measurements. Energy dispersive X-ray spectroscopy (EDS) was performed in a scanning electron microscope (SEM, FEI Quanta 250) with a silicon drift detector attachment (Oxford X-Max). EDS was performed to measure the composition of the FGM, using both line scan and area mapping techniques. X-ray diffraction (XRD) was used for phase identification within the FGM, and the diffraction patterns were collected using a Bragg-Brentano-type diffractometer (Panalytical Empyrean, Cu K-α X-ray source, 45 kV, 40 mA, λ=1.54 Å). Electron backscatter diffraction (EBSD, Oxford Nordlys Max2) was performed on selected regions of the sample to analyze local phase compositions in areas of interest. Microhardness was measured along the height of the FGM (Leco LM 110AT Hardness Tester).

Tensile samples, with the geometry shown in Figure 1c, were extracted by wire EDM from the half of the pillar not used for metallographic characterization. Samples were extracted from two vertical locations. A total of ten tensile samples were extracted: five with the gauge region containing between 0 vol % and 30 vol% NiCr (bal. SS304L) and five with the gauge region containing between 60 vol % and 90 vol% NiCr (bal. SS304L) as shown in Figure 1b. These will be referred to as the SS304L-rich and NiCr-rich tensile samples, respectively.

Uniaxial tensile tests were performed using a custom-built mini test stage with a load limit of 4500 N. Samples were loaded with an approximate strain rate of $1.5 \times 10^{-4}$ s$^{-1}$. During deformation, images were taken at 1 Hz using a digital camera (Point Grey GRAS-50S5M-C), and digital image correlation (DIC), a non-contact surface deformation measurement technique, was used to measure the evolving in-plane surface deformation fields. The deformation fields were computed using DIC analysis software (Vic-2D 6, Correlated Solutions) with a cubic B-spline interpolation algorithm with a subset size of 29 pixels and a step size of 7 pixels, with a pixel size of 15 µm. A 19 mm long vertical virtual extensometer was used to compute an effective strain.

*2.2. Computational techniques*

Thermodynamic analysis of the NiCr/SS304L system was performed to predict the phase relations within the gradient. These calculations were performed in terms of the CALPHAD method, in which the Gibbs energy of individual phases are parameterized with the multicomponent compound energy formalism in the form of sublattice models [20–22]. This analysis was performed using pycalphad [23] and Thermo-Calc [24] using a Cr-Fe-Ni database developed by Hallstedt [25]. Scheil-Gulliver simulations (also referred to as Scheil simulations) were performed in order to model the solid phases that precipitate from the melt during solidification. Scheil simulations assume that the liquid phase is well-mixed and homogenous in composition as the melt solidifies, diffusion in the solid is negligible, and that there is local equilibrium at the solid/liquid interface. These assumptions are valid for modeling solidification during the AM process because of the rapid solidification and the melt pool being well-mixed due to Marangoni flow [26]. Scheil simulations in AM FGMs have been shown to better predict the phase composition compared to equilibrium calculations as they can be considered to represent the upper bound of solute partitioning between the liquid and solid phases [27–29] and experimental validation has shown that the rapid cooling and melt behavior are well approximated by this model. The *scheil* module of pycalphad, developed by this team, was used for these simulations [29,30].

To computationally investigate the tensile behavior of the NiCr/SS304L FGM, models of the two sample types were generated using the finite element analysis (FEA) software ABAQUS. The tensile sample geometry was meshed with 10,020 elements with a minimum size of 150 µm. The yield stress for each gradient composition was initially estimated using the measured Vickers microhardness. The final constitutive models for each region within these samples, including yield strength and strain hardening behavior, was determined through an iterative approach of refining the elasto-plastic input parameters of the various compositions within the simulation and comparison with the experimentally measured force-displacement behavior as well, with a secondary comparison of the FEA strain fields with those measured experimentally using DIC.

## 3. Results and Discussion

### 3.1. Composition analysis

In order to confirm the deposited composition, EDS line scan analysis was performed along the entire sample. The results indicated no large deviations between the measured composition and the planned deposition composition, and there were clear steps between the gradient regions, as shown in Figure 2. The regions marked by lighter and darker shades of gray indicate the composition of the gauge regions of the tensile specimen. The NiCr-rich sample was characterized by larger fluctuations in elemental compositions, which is an indication of elemental heterogeneity between disparate phases as discussed further in Section 3.4.

*3.2. Characterization of mechanical behavior*

Vickers microhardness measurements were taken in all eleven of the gradient composition regions of the FGM, with the results shown in Figure 3. These data show a monotonically increasing hardness from the 0 vol% NiCr/100 vol% SS304L region, with a hardness of 223 HV, to the 60 vol% NiCr/40 vol% SS304L region, with a hardness of 445 HV, at which point the hardness plateaus at an average value of 441 HV. Within the regions contained in the tensile samples, the lowest hardness in the SS304L-rich tensile sample was 220 HV (10 vol% NiCr/90 vol% SS304L), and the lowest hardness in the NiCr-rich tensile sample was 429 HV (90 vol% NiCr/10 vol% SS304L).

Engineering stress-strain curves for both composition spans are shown in Figure 4. The average ultimate tensile strength, yield strength, and elongation to failure for the two sample types are given in Table 1. The SS304L-rich tensile samples exhibited ductile behavior while the NiCr-rich tensile samples failed after negligible plastic deformation. Optical microscope and SEM images of the fracture surfaces of these samples are given in Figure 5.

*3.3. Finite element analysis simulations*

FEA simulations were performed to identify the constitutive behavior of different composition regions within the FGMs using an iterative method. The elastic modulus of each gradient region was estimated using a rule of mixtures of the volume fractions and elastic moduli of the two terminal materials, which were taken to be 193 GPa for SS304L [31] and 220 GPa for NiCr [19]. The yield strength of each region was initially estimated from the measured Vickers microhardness in that region using the equation,

$$YS = -90.7 + 2.876\, H_v \tag{1}$$

where *YS* is the yield strength and $H_v$ is the Vickers microhardness [32]. This relationship was empirically derived by Pavlina et al. using yield strength and hardness data for steels with tensile strength to yield strength ratios that ranged from less than 1.23 to greater than 1.56 in order to account for different strain hardening behaviors [32]. Here, this relationship is applied to AM NiCr, which has a tensile strength to yield strength ratio of 1.55. For each sample type, the initial estimate of the strain hardening slope was approximated to be the same for each composition within the sample, and equivalent to that of the experimentally measured macroscopic engineering stress-strain behavior.

Through an iterative process, the yield strength for each region was modified to improve the agreement between the simulated and experimentally measured engineering stress-strain behavior as well as the evolution of strain fields within the gauge region of the samples. The final flow stress versus plastic strain curves used in the FEA simulations are given in Figure 6. The dashed lines in both plots correspond to the maximum engineering stress reached by each sample type during deformation, differentiating the regions that did or did not plastically deform during macroscopic deformation of the FGM samples.

Comparisons between experimentally obtained and simulated stress-strain behavior and strain fields are given in Figure 7. For the SS304L-rich tensile sample, both the experimentally measured (Figure 7a) and simulated (Figure 7b) strain fields show a strong gradient between the different composition regions within the gauge section of the tensile specimen. The strain localized at the interface between the 0 vol% NiCr/100 vol% SS304L region and the 10 vol% NiCr/90 vol% SS304L region in both experiments and simulations, which is where these samples failed. The simulated engineering stress-strain curve agreed well with experimental results as shown in Figure 7c.

Similarly, the NiCr-rich sample showed good agreement between experimentally measured and simulated results, in terms of the strain fields immediately prior to failure (Figure 7d and e) and engineering stress-strain curves (Figure 7f). The experimentally measured and simulated strain fields are shown in Figure 7d and e with small uniform strains (under 0.0075) throughout the gauge region and localization at the interface between the 80 vol% NiCr/20 vol% SS304L and 90 vol% NiCr/10 vol% SS304L composition regions. Failure occurred in the 90 vol% NiCr region of the sample.

*3.4. Experimental characterization of phase composition in failure regions*

XRD, EDS, and EBSD analysis were performed on the two gradient regions of interest where failure occurred in the two different tensile specimen types: the 10 vol% NiCr/90 vol% SS304L and 90 vol% NiCr/10 vol% SS304L gradient regions. X-ray diffraction patterns for both composition regions are given in Figure 8. The diffraction pattern of the 10 vol% NiCr/90 vol% SS304L region indicates the presence of Fe-rich FCC peaks and an overlapping Cr-rich BCC phase peak at 44°, which agrees with computational phase analysis to be shown in Section 3.5. The XRD pattern of the 90 vol% NiCr/10 vol% SS304L region include Ni-rich FCC phase peaks, with the same overlapping Cr-rich BCC phase peak at 44°.

EDS maps of the 10 vol% NiCr/90 vol% SS304L composition region (Figure 9) show an even distribution of Ni throughout the entire region with areas of Fe and Cr elemental segregation. The overall elemental composition of this gradient region is 63 wt% Fe, 17 wt% Ni, and 20 wt% Cr. The Fe-rich cellular regions contain on average about 2 wt% more Fe and 1 wt% less Cr than the overall gradient region composition, while the Cr-rich inter-cellular regions contain about 7 wt% more Cr, 2 wt% less Ni and 5 wt% less Fe than the overall layer composition. The 90 vol% NiCr/10 vol% SS304L composition region had a dendritic structure with elemental segregation, as shown in Figure 9f. While Fe is present throughout, the concentration of Fe is about 3 wt% higher in the inter-dendritic regions than the dendrites, as evidenced by the Fe EDS map in Figure 9g. The dendrites are Cr-rich, and the inter-dendritic regions are Ni rich, although both elements are present throughout the entire area (Figure 9h and i). The average composition of the Cr-rich dendrites measured using EDS analysis was 63 wt% Cr, 32 wt% Ni, and 5 wt% Fe.

Localized EBSD phase analysis of the same areas of the SS304L-rich and NiCr-rich tensile samples showed that the 10 vol% NiCr/90 vol% SS304L composition region was composed of Fe-rich FCC phase in the cellular region of the microstructure, as shown in Figure 9e, with the phase fraction data given in Table 2. This agrees with the XRD data from this same area. Note that the 30 area% zero solution, or area that could not be resolved by the EBSD software, was primarily located in the inter-cellular region. This region was Cr-rich and likely composed of the Cr-rich BCC phase. Image analysis of the Cr EDS map indicated that the Cr-rich area accounted for approximately 11% of the total image area. This could be considered as the area that would be the BCC phase and thus the true zero solution is only about 19%. Volume changes during solidification could result in large strains and high dislocation densities in this region, making phase identification in this small area difficult.

The 90 vol% NiCr/10 vol% SS304L region was composed of a two-phase NiCr-rich FCC and Cr-rich BCC eutectic composition as shown in Figure 9j. The FCC phase is the majority phase (72.5 area% of the analyzed area), and the BCC phase occupies only 2.19 area% of the analyzed area, while 25.1 area% could not be resolved by the EBSD analysis software. Similar to the 10 vol% NiCr/90 vol% SS304L region, this region was also characterized by Cr-rich areas in the EDS composition maps that correspond to the BCC phase area identified via EBSD. Further image analysis of the Cr EDS map shown in Figure 9i indicated that the Cr-rich regions consisted of 17 area% of the total analyzed area. Therefore, it is hypothesized that the 17 area% would also be the Cr-rich BCC phase, for a total of 19.19 area% of Cr-rich BCC phase and only 8.1 area% of true zero solution. A majority of the zero solution is concentrated in this area due to either difficulties in achieving an even polish in the two-phase regions, or a high concentration of strain and dislocations.

Previous studies of Ni-Cr alloys have shown that the eutectic composition is brittle with a tensile ductility ranging from 1-2.5% [33]. Thus, it is hypothesized that the reason for the low ductility of the NiCr-rich samples was due to low ductility of NiCr further hindered by low ductility of the inter-cellular eutectic phase. The experimental results from the mechanical tests, composition analysis, and phase analysis show how both the inherent mechanical properties of the two constituent alloys and the phase formation that results from combining SS304L and NiCr in an FGM play a role in the mechanical properties of the full gradient. The results highlight that the overall ductile behavior of SS304L and brittle behavior of NiCr are preserved in the SS304L/NiCr FGM but the formation of the brittle eutectic phase further reduced the ductility within the regions where it formed (90 vol% NiCr/10 vol% SS304L).

*3.5. Thermodynamic calculation results*

Computational phase analysis was performed along the full composition range of the FGM. Figure 10 plots the equilibrium isopleth along the gradient path, showing that the temperature difference between the liquidus/solidus lines is extremely small across the entire gradient region. This indicates that the temperature window where solidification occurs is also very small. The solidus temperature at each composition was used to calculate the equilibrium phase composition for the corresponding composition along the entire FGM as shown in Figure 11a, together with the phase composition as calculated using Scheil simulations in Figure 11b. Both calculation methods indicated a shift from an FCC+BCC mixture to an FCC-only composition from SS304L, represented by Fe-18Cr-8Ni, with the primary difference being the composition where this transition occurs. The equilibrium calculation indicates this transition will occur at 20 vol% NiCr/80 vol% SS304L, and Scheil simulation indicates this will occur at 30 vol% NiCr/70 vol% SS304L; experimentally, both the FCC and BCC phases were observed throughout the entire gradient.

The liquidus projection of the Cr-Fe-Ni system, Figure 12, shows the primary crystallization phases - BCC on the Cr/Fe-rich side and FCC on the Ni-rich side - and the monovariant line indicating the compositions where both of these phases have zero amount and are in equilibrium with the liquid. The monovariant line connects the peritectic reaction occurring in the Fe-Ni binary system at 1517°C to the eutectic reaction occurring in the Cr-Ni binary system at 1346°C. Figure 12 has two superimposed broken lines, with the horizontal dotted line corresponding to the isopleth in Figure 10, marking the gradient path from SS304L to NiCr, and the inclined line with dotted and dashed portions corresponding to the isopleth in Figure 13.

The intersection of the SS304L to Ni-20Cr line and the monovariant line in Figure 12 indicates the composition where the primary crystallization phase changes from BCC to FCC. With increasing NiCr fraction, crossing the monovariant line means that FCC will be the first phase to form, but BCC may still form in both equilibrium and Scheil solidification simulations. In equilibrium solidification simulations, the presence of BCC for NiCr-rich compositions depends only on the phases present at the solidus. In Scheil solidification simulations, the segregation of solute from the melt will push the liquid composition towards the monovariant line and then the liquid composition will follow the monovariant line, forming both BCC and FCC, until it reaches the eutectic or until the liquid is completely solidified. The outer points on the inclined line denote the composition ends of the isopleth in Figure 13 which contains the Scheil solidification path for the 90 vol% NiCr/10 vol% SS304L gradient region. The inner points correspond to the compositions of the first and last FCC solids to form, marked by the two vertical lines in Figure 13, demonstrating that the segregation of the liquid at NiCr-rich portions of the gradient do not reach the monovariant line before the liquid is fully solidified in the Scheil simulation.

In the 10 vol% NiCr/90 vol% SS304L region, the amounts of the BCC phase in the equilibrium calculation and Scheil simulations are 0% and 8.3%, as given in Table 3. In both simulations, the FCC phase is predicted to be the first phase to solidify, followed by the BCC phase. The 11 area% zero solution attributed to the BCC phase in the ESBD phase map of the region is in semi-quantitative agreement with the thermodynamic calculations.

For the 90 vol% NiCr/10 vol% SS304L region, both equilibrium calculations and Scheil simulations predicted the formation of the FCC phase only as shown in Figure 11a, while EBSD phase analysis indicated the presence of the BCC phase in the inter-dendritic region of the gradient region as shown above. The compositions of the FCC phase that formed during the Scheil solidification simulation are 8 wt% Fe, 78 wt% Ni, and 15 wt% Cr at the beginning and 6 wt% Fe, 32 wt% Ni, and 62 wt% Cr at the end. The isopleth between these two compositions, given in Figure 13, shows that the Scheil solidification-driven micro-segregation results in final liquid/solid compositions that are far from the nominal 90 vol% NiCr/10 vol% SS304L composition. It can be seen that the FCC phase can transform to the mixture of FCC+BCC at relatively high temperatures, with the zero phase fraction line for the BCC phase rapidly increasing in temperature as the composition approaches the composition of the monovariant line that this isopleth cuts through. If the last solid to form were to cool under equilibrium conditions, BCC would begin to form below 1106°C and would reach a phase fraction of 30% at 400°C. The first BCC to form would have a Cr-enriched phase composition of 6 wt% Fe, 51 wt% Ni and 43 wt% Cr, which is comparable to the composition of the Cr-rich dendrites measured by EDS of 63 wt% Cr, 32 wt% Ni, and 5 wt% Fe (Figure 9j). The presence of BCC in the equilibrium isopleth along the solidification path of 90 vol% NiCr/10 vol% SS304L and the phase composition of the first BCC to form supports the identification of BCC in the EBSD analysis. These calculations further validate the use of Scheil solidification simulations to predict the behavior of additively manufactured materials because the experimentally observed BCC phase would be unlikely to form in NiCr-rich gradient regions without solidification-driven solute segregation to produce solid phases with compositions that favor BCC phase stability.

## 4. Conclusions

In this study, the phase composition, hardness, and tensile behavior across gradient compositions of a SS304L to NiCr FGM made by DED AM were analyzed through both experimental characterization and computational simulations. Two sample types were analyzed, one that was SS304L-rich and another that was NiCr-rich, and the primary findings of the study are as follows:

- The SS304L-rich tensile specimens, spanning compositions of 0 vol % and 30 vol% NiCr (bal. SS304L), experienced ductile fracture at a composition of 10 vol% NiCr/90 vol% SS304L, where FCC and BCC phases were predicted computationally and found experimentally with quantitative agreement.
- Conversely, the NiCr-rich tensile samples, spanning compositions of 60 vol % and 90 vol% NiCr (bal. SS304L), experienced brittle fracture after negligible macroscale plastic deformation. Fracture occurred at a composition of 90 vol% NiCr/10 vol% SS304L where FCC and BCC eutectic phases were found experimentally and predicted computationally by considering the segregation due to Scheil simulations.
- Elemental mapping of the 10 vol% NiCr/90 vol% SS304L sample showed a cellular structure with a slightly Fe-rich cellular region and slightly Cr-rich inter-cellular region. The same analysis of the 90 vol% NiCr/10 vol% SS304L region showed strong Cr-rich elemental segregation in the inter-dendritic region of the microstructure and an Fe-rich dendritic structure. Thermodynamic analysis of the 90 vol% NiCr/10 vol% SS304L region supports the formation of inter-dendritic Cr-rich BCC through solid state transformations at relatively high temperatures from FCC formed through Scheil solidification-driven segregation of the liquid.
- Phase analysis of these regions indicated primarily FCC phase composition for both regions. In the 90 vol% NiCr/10 vol% SS304L region, the BCC phase was present in the Cr-rich inter-dendritic region. Scheil simulations did not indicate the presence of BCC phase in the 90 vol% NiCr/10 vol% SS304L, but closer examination of the solidification path between the first and last solids to form in this region shows that solidification-driven segregation can produce solid compositions that favor a solid-state transformation to BCC at high temperatures, where the last solid to form can begin to form BCC at 1106˚C.

- No deleterious intermetallic phases were predicted to form through solidification or high-temperature solid state transformations. It was determined that the inherently brittle nature of NiCr, along with the formation of a brittle eutectic phase, resulted in macroscale brittle failure behavior of the NiCr-rich sample. Conversely, the ductile nature of SS304L with the absence of deleterious secondary phases resulted in a ductile behavior in the SS304L-rich sample.

**Acknowledgments**

Part of this research was carried out at the Jet Propulsion Laboratory, California Institute of Technology, under a contract with the National Aeronautics and Space Administration. LDB was supported by an NDSEG Fellowship. BB was supported by a NASA Space Technology Research Fellowship (grant 90NSSC18K1168) and an NSF National Research Trainee Fellowship (grant DGE-1449785).

**Figures**

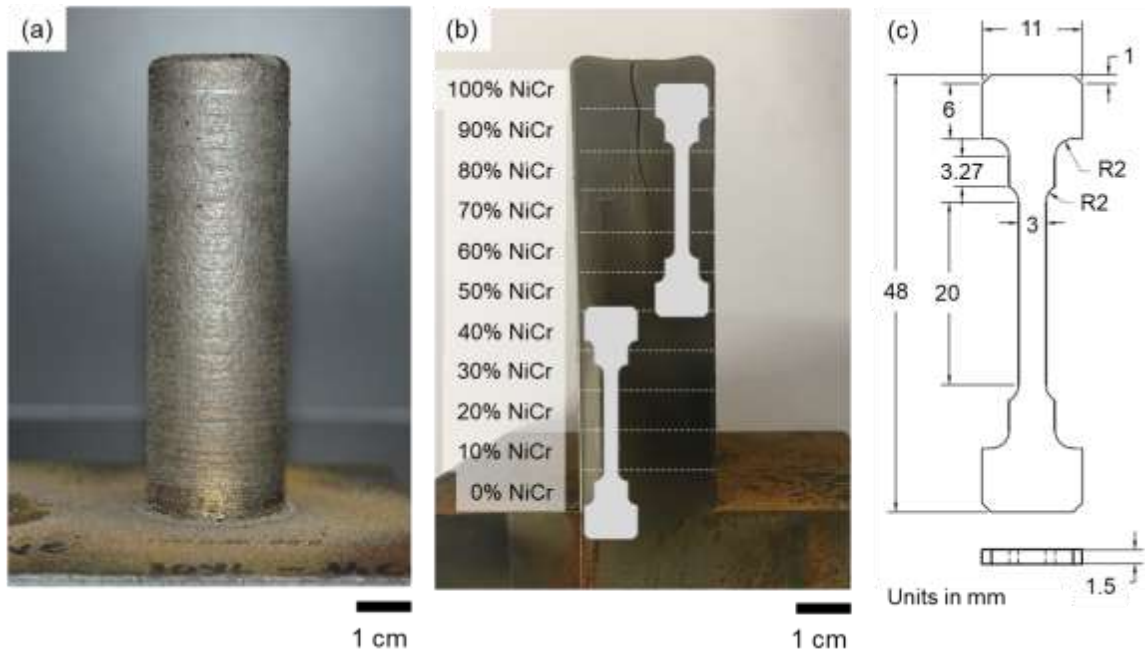

**Figure 1.** (a) Photograph of the as-built SS304L/NiCr FGM sample (a) prior to sectioning and (b) after sectioning, where the composition changes are marked by dashed white lines with compositions noted, and the locations from which the two types of tensile samples (SS304L-rich and NiCr-rich) were extracted are schematically shown. (c) Geometry of the tensile specimens, with all measurements in mm.

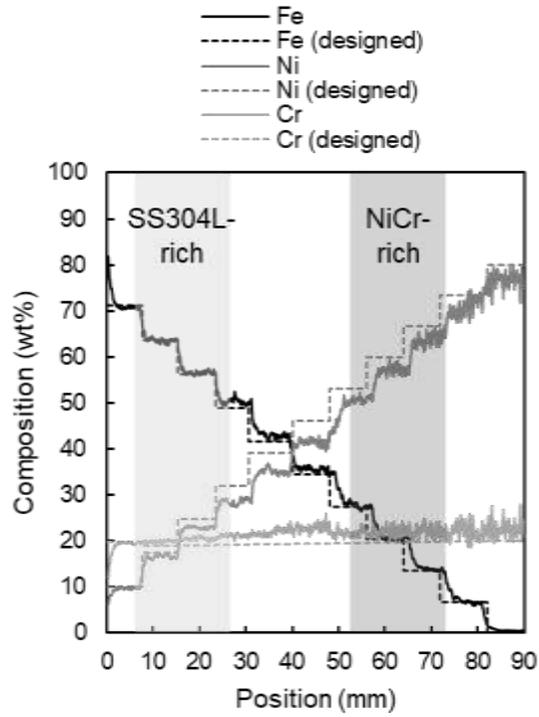

**Figure 2.** Composition as a function of position along the build direction of the SS304L/NiCr FGM. The solid lines indicate the composition as measured by EDS analysis, while the dashed lines indicated the planned composition. The light grey shaded region corresponds to the portion of the FGM included in the SS304L-rich tensile specimens and the dark grey shaded region corresponds to the portion of the FGM included in the NiCr-rich tensile specimens.

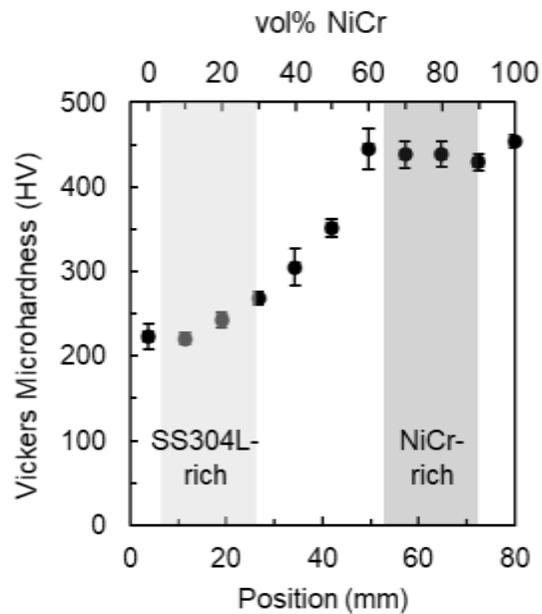

**Figure 3.** Vickers microhardness as a function of distance from the baseplate in the full SS304L/NiCr FGM. Each symbol corresponds to the average hardness from one region with the error bars indicating standard deviation for the five measurements within that region. The light grey shading corresponds to the portion of the FGM included in the SS304L-rich tensile specimens and the dark grey shading corresponds to the portion of the FGM included in the NiCr-rich tensile specimens.

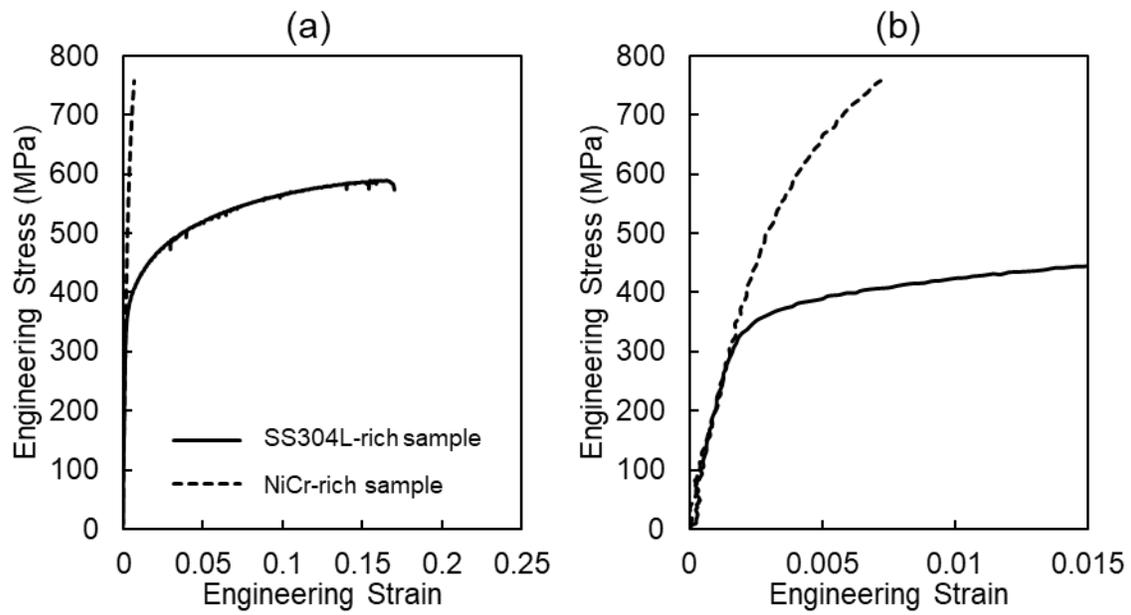

**Figure 4.** Engineering stress-strain data for (a) representative SS304L-rich and NiCr-rich samples with (b) a zoom-in of the small strain region.

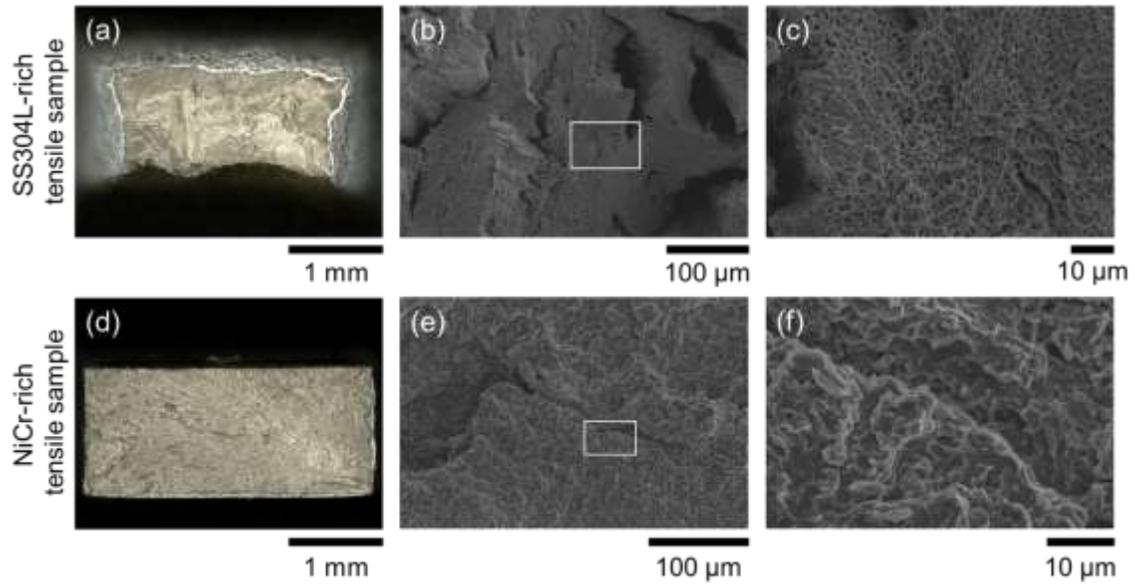

**Figure 5.** Optical and scanning electron microscope image of the fracture surfaces of a representative sample of (a-c) the SS304L-rich tensile sample and (d-f) the NiCr-rich tensile sample highlighting the ductile fracture of the SS304L-rich samples and brittle fracture character of the NiCr-rich samples.

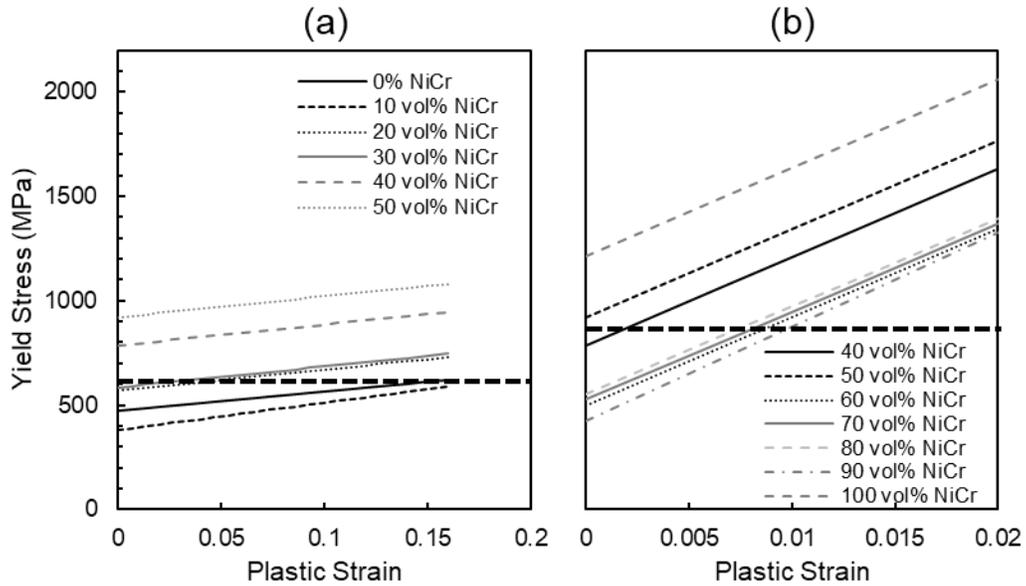

**Figure 6.** Yield stress versus plastic strain used as input in FEA simulations of the (a) SS304L-rich sample and (b) NiCr-rich sample to produce the experimentally observed stress-strain curves and surface strain contours for each sample type. The horizontal dashed lines indicate the max engineering stress reached by each sample type during tensile tests, delineating which regions deformed plastically (below the dashed line) and those that did not yield (above the dashed line).

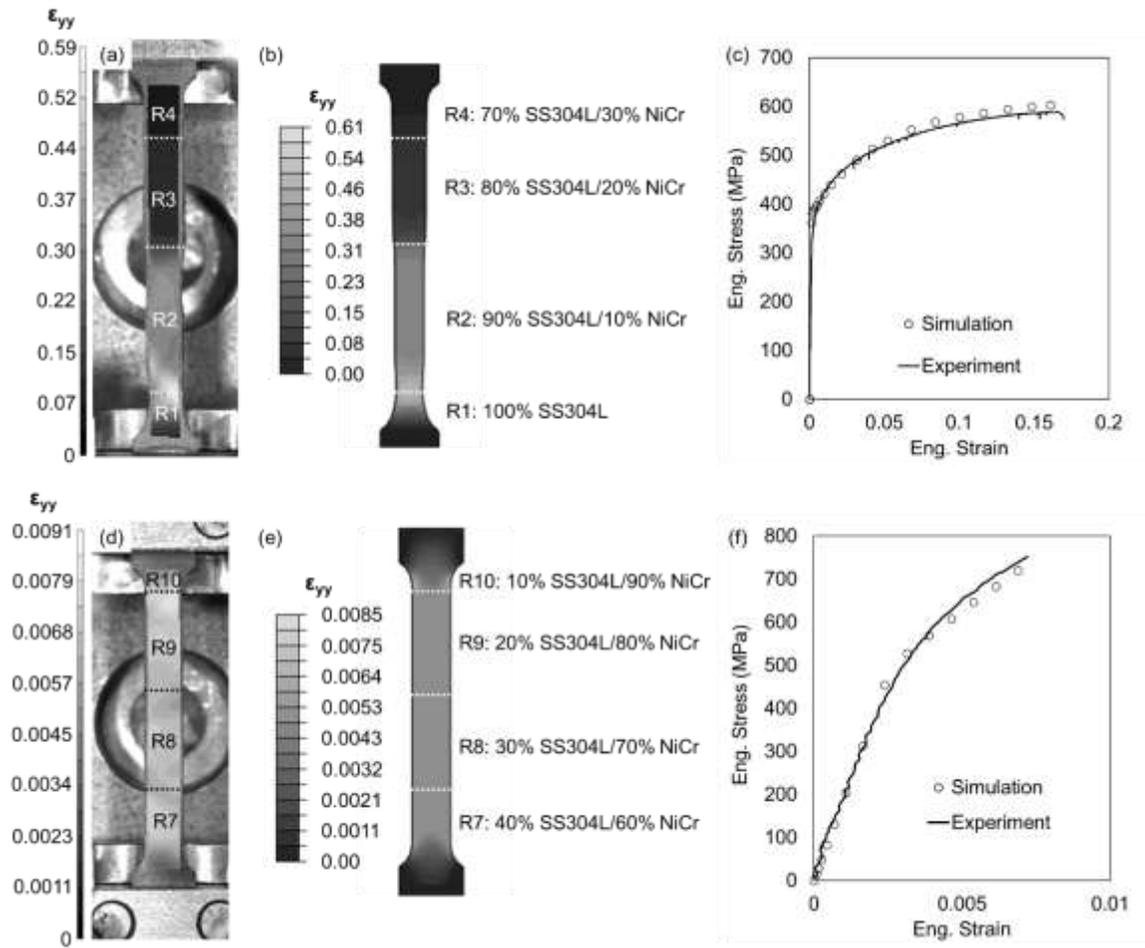

**Figure 7.** Comparison of experiments and finite elements simulations of the SS304L-rich sample (top row) and NiCr-rich sample (bottom row). Comparison of strain fields via (a, d) experimentally measured DIC immediately prior to fracture and (b, e) the FEA simulation at an applied displacement corresponding to experimentally measured displacement to failure. (c, f) Engineering stress-strain curves from a representative experiment (line) versus the corresponding FEA simulation (symbols).

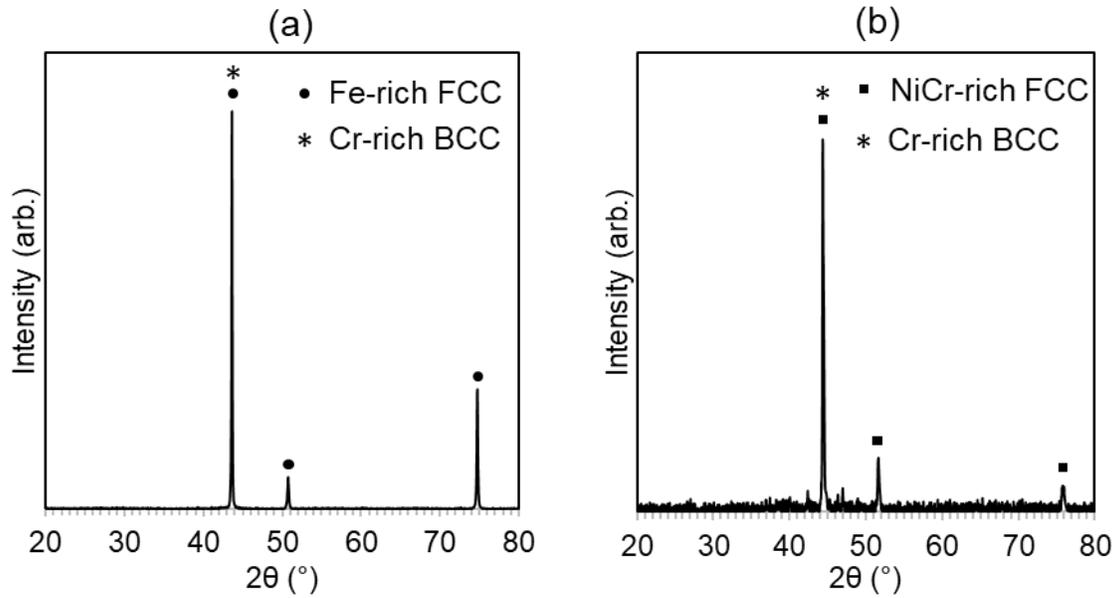

**Figure 8.** X-Ray diffraction patterns from (a) 10 vol% NiCr/90 vol% SS304L (SS304L-rich) and (b) 90 vol% NiCr/10 vol% SS304L (NiCr-rich) regions of the FGM. The Fe-rich FCC, Cr-rich BCC, and Ni-rich FCC phases are identified and labeled.

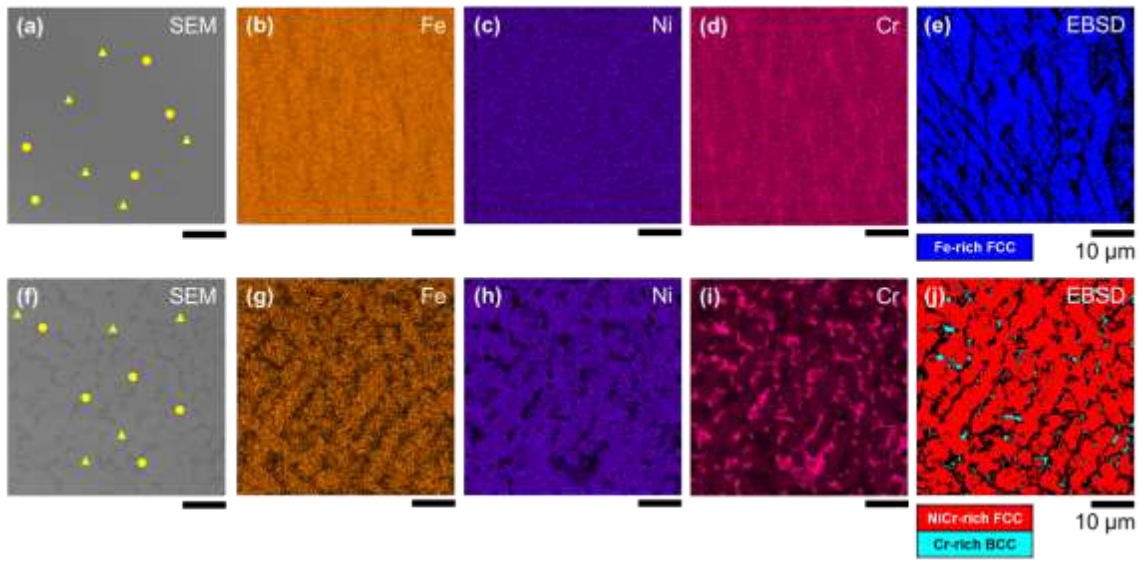

**Figure 9.** Results from characterization of selected regions of the SS304L-rich sample (top row) and NiCr-rich sample (bottom row), including (a, f) SEM images and (b-d, g-h) EDS composition maps of the main constituent elements, (b, g) Fe, (c, h) Ni, and (d, i) Cr, and (e, j) EBSD phase maps of the areas shown in the SEM and EDS images. The markers on the SEM image indicate locations where point composition analysis was performed in the FCC (triangle) and BCC (circle) regions.

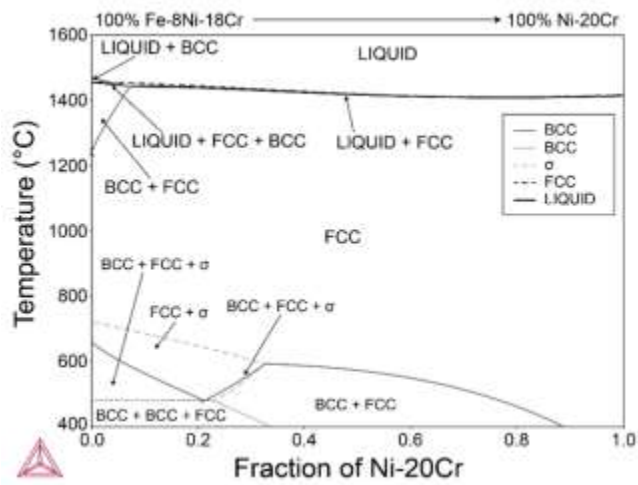

**Figure 10.** Isopleth along the SS304L to NiCr gradient path between 400°C and 1600°C.

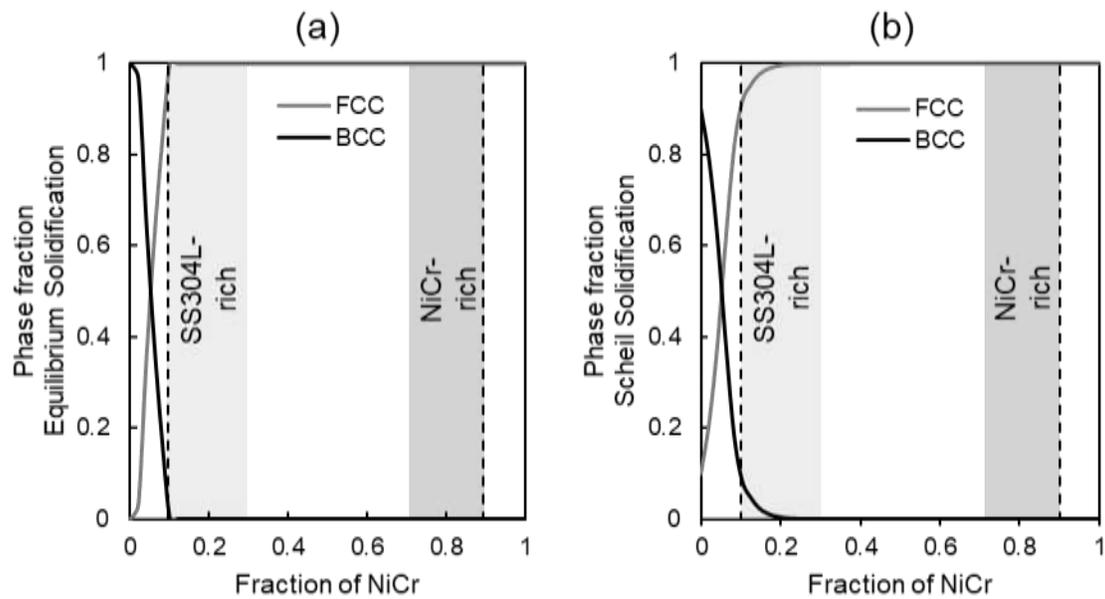

**Figure 11.** (a) Phase fraction of the BCC and FCC phases as a function of fraction of NiCr, as calculated at equilibrium at the solidus temperature for each composition and (b) using Scheil-Gulliver simulations. The light and dark grey shaded regions correspond to the compositions of the SS304L-rich and NiCr-rich tensile samples, respectively. The dashed vertical lines on both graphs correspond to the compositions where the two different sample types failed.

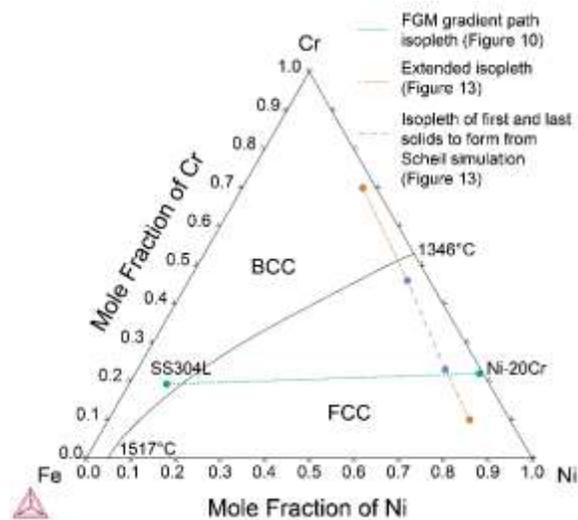

**Figure 12.** Liquidus projection of Cr-Fe-Ni showing the gradient path from SS304L to NiCr that corresponds to the isopleth in Figure 10 (horizontal dotted line). The inclined dashed and dotted line shows the path intersecting the first solid to form and the last solid to form on the solidification path corresponding to the 90 vol% NiCr/10 vol% SS304L gradient region, shown as an isopleth in Figure 13.

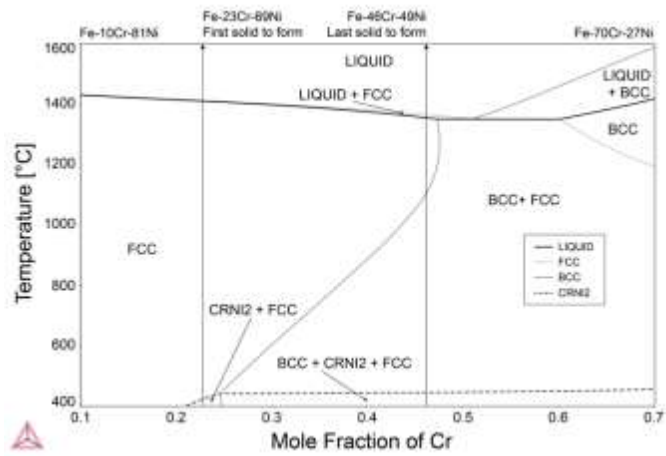

**Figure 13.** Equilibrium isopleth along the path from the first solid to form to the last solid to form in the Scheil solidification simulation for the 90 vol% NiCr/10 vol% SS304L gradient region.

**Tables**

**Table 1.** Average ultimate tensile strength (UTS), yield strength (YS) and % elongation for the SS304L-rich and NiCr-rich samples.

| Sample | Avg. UTS (MPa) | Avg. YS (MPa) | Avg. % Elongation |
|---|---|---|---|
| SS304L-rich | 591 ± 7.8 | 390 ± 7.9 | 18 ± 2.7 |
| NiCr-rich | 825 ± 51 | 727 ± 19 | 0.85 ± 0.2 |

**Table 2.** EBSD phase data for the 10 vol% NiCr/90 vol% SS304L (SS304L-rich) and 90 vol% NiCr/10 vol% SS304L (NiCr-rich) gradient regions of the tensile samples. All values are in area% and dashes indicate that the phase is not present in the given region.

| Phase Name | 10 vol% NiCr | 90 vol% NiCr |
|---|---|---|
| Fe-rich FCC | 69.8 | - |
| NiCr-rich FCC | - | 72.5 |
| Cr-rich BCC | - | 2.2 |
| Zero Solution | 30.2 | 25.1 |

**Table 3.** Phase fraction data as calculated via equilibrium and Scheil solidification for the 10 vol% NiCr/90 vol% SS304L and 90 vol% NiCr/10 vol% SS304L gradient regions.

| Phase Name | 10 vol% NiCr | | 90 vol% NiCr | |
|:---:|:---:|:---:|:---:|:---:|
| | Equilibrium | Scheil | Equilibrium | Scheil |
| FCC | 0.958 | 0.876 | 1.0 | 1.0 |
| BCC | 0.042 | 0.124 | 0.0 | 0.0 |